\documentclass[pra,11pt,superscriptaddress]{revtex4}
\usepackage{times,calc}
\usepackage{amsbsy,amssymb}
\usepackage{graphicx,color}

\newcommand{\ignore}[1]{}

\ignore{
latex ndgates ; dvipsa ndgates

tar cvf ndgates.tar ndgates.tex ndgates.bbl cs2_27.eps
gzip ndgates.tar
}

\newcommand{\qvbar}{\mbox{$|\hspace*{-3pt}|\hspace*{-3pt}|$}}
\newcommand{\qrangle}{\mbox{$\rangle\hspace*{-4.3pt}\rangle\hspace*{-4.3pt}\rangle$}}
\newcommand{\qlangle}{\mbox{$\langle\hspace*{-4.3pt}\langle\hspace*{-4.3pt}\langle$}}

\newcommand{\sysfnt}{\mathsf}

\newcommand{\ket}[1]{\qvbar{#1}\qrangle}

\newcommand{\kets}[2]{\qvbar{#1}\qrangle_{{}_{\!\!{\scriptstyle\sysfnt{#2}}}}}
\newcommand{\bras}[2]{{}^{\scriptstyle\sysfnt{ #2}}\!\qlangle{#1}\qvbar}

\newcommand{\slb}[2]{{#1}^{({\sysfnt{#2}})}}

\newcommand{\bop}[1]{{\mathbf{#1}}}
\newcommand{\killops}[1]{\slb{\killop}{#1}}

\newcommand{\killop}{\bop{a}}

\newcommand{\crtops}[1]{\slb{\bop{c}}{#1}}

\newcommand{\vacuum}{\ket{\bop{0}}}
\newcommand{\vacuums}[1]{\kets{\bop{0}}{#1}}

\newlength{\elimdepthdim}
\newlength{\elimheightdim}
\newlength{\elimwidthdim}

\newlength{\strutdepthdim}
\newlength{\strutheightdim}
\newlength{\strutwidthdim}

\newcommand{\bitzero}{\mathfrak{0}}
\newcommand{\bitone}{\mathfrak{1}}

\newcommand{\idop}{\textbf{I}}

\begin{document}

\title{A Note on Linear Optics Gates by Post-Selection}
\author{E. Knill}
\affiliation{MS B265, Los Alamos National Laboratory, Los Alamos, NM 87545}
\email{knill@lanl.gov}

\date{\today}

\begin{abstract}
Recently it was realized~\cite{knill:qc2000e} that linear optics and
photo-detectors with feedback can be used for theoretically efficient
quantum information processing.  The first of three steps toward
efficient linear optics quantum computation (eLOQC) was to design a
simple non-deterministic gate, which upon post-selection based on a
measurement result implements a non-linear phase shift on one mode.
Here a computational strategy is given for finding non-deterministic
gates for bosonic qubits with helper photons. A more efficient
conditional sign flip gate is obtained.
\end{abstract}

\maketitle

\section{Introduction}

Now that we know that linear optics and photo-detectors are sufficient
for quantum information processing~\cite{knill:qc2000e}, it is
necessary to investigate how the necessary schemes can be realized
more efficiently. One promising direction is to use superpositions of
squeezed or coherent states for encoding
qubits~\cite{gottesman:qc2000a,ralph:qc2002b}.  In this note, it is
shown how the non-deterministic gates at the foundation of the
constructions in~\cite{knill:qc2000e} can be found and improved.
Other relevant work in this direction includes
\cite{ralph:qc2001a,rudolph:qc2001a,pittman:qc2001a,pittman:qc2001b},
where networks suitable for experimental realization are given.  The
focus of this paper is on what can be done in principle without giving
experimentally accessible layouts.  To that end, a systematic method
is given for finding non-deterministic gates based on a combination of
algebraic solution finding, exploitation of known symmetries, and
numerical optimization.  By using the method, a conditional sign flip
for bosonic qubits that succeeds with probability $1/13.5$ using two
helper photons is found. This improves the one
in~\cite{knill:qc2000e}, which succeeds with probability $1/16$. What
is the optimum probability of success for any number of helper
photons?  A characterization of the achievable states without
post-selection implies that the probability cannot be one, a result
related to known bounds on
Bell-measurements~\cite{lutkenhaus:qc1999a,calsamiglia:qc2001a}.

\section{Preliminaries}

The physical system of interest consists of bosonic modes,
each of whose state space is spanned by the number states
$\ket{0},\ket{1},\ket{2},\ldots$. If more than one mode is used, they
are distinguished by labels. For example, $\kets{k}{r}$ is the state
with $k$ photons in the mode labeled $\sysfnt{r}$.  The 
hermitian transpose of this state is denoted by $\bras{k}{r}$. The
vacuum state for a set of modes has each mode in the state $\ket{0}$
and is denoted by $\vacuum$.  The anniliation operator for mode $r$ is
written as $\killops{r}$ and the creation operator as
$\crtops{r}=\left(\killops{r}\right)^\dagger$. Recall that
$\crtops{r}\ket{m}=\sqrt{m+1}\ket{m+1}$.  Labels are omitted when no
ambiguity results. Hamiltonians that are at most quadratic in creation
and annihilation operators generate the group of linear optics
transformations.  Among these, the ones that preserve the particle
number are called passive linear. Every passive linear optics
transformation can be achieved by a combination of beam splitters and
phase shifters. If $U$ is passive linear, then $U\crtops{r}\vacuum =
\sum_{s}u_{sr}\crtops{s}\vacuum$, where $u_{sr}$ defines a unitary
matrix $\hat u$.  Conversely, for every unitary matrix $\hat u$, there is a
corresponding passive linear optics transformation~\cite{reck:qc1994a}. 
For the remainder of this note, all linear optics transformations
are assumed to be passive.

\section{Conditional phase shifts}

A conditional phase shift by $\theta$ on two modes is the map
$\mathsf{CS}_{\theta}:\ket{ab}\rightarrow e^{i (ab)\theta}\ket{ab}$
for $0\leq a,b\leq 1$. These phase shifts can be used to
implement conditional phases on two bosonic qubits.  A bosonic qubit
$\mathsf{Q}_{\sysfnt{r},\sysfnt{s}}$ is defined by identifying logical
$\ket{\bitzero}$ with $\kets{01}{rs}$ and logical $\ket{\bitone}$ with
$\kets{10}{rs}$. The modes $\sysfnt{r}$ and $\sysfnt{s}$ can be two
distinct spatial modes or the two polarizations of one spatial mode.
To realize the conditional sign flip between
$\mathsf{Q}_{\sysfnt{1},\sysfnt{2}}$ and
$\mathsf{Q}_{\sysfnt{3},\sysfnt{4}}$, apply $\mathsf{CS}_{180^\circ}$
to modes $\sysfnt{1}$ and $\sysfnt{3}$.  The bosonic qubit
controlled-not can then be implemented using conditional sign flips
and single qubit rotations, which are realizable with beamsplitters.

In~\cite{knill:qc2000e}, conditional sign flips were implemented
indirectly using a non-deterministic realization of the map
\begin{equation}
\mathsf{NS}:\alpha\ket{0}+\beta\ket{1}+\gamma \ket{2}\rightarrow \alpha\ket{0}+\beta\ket{1}-\gamma \ket{2}
\end{equation}
that succeeds with probability $1/4$.  This realization requires one
helper photon and two ancilla modes.  The goal is to implement
$\mathsf{CS}_{\theta}$ more efficiently directly using two helper
photons.  One helper photon can be shown to be insufficient by means
of the same algebraic method about to be used.  Let modes $\sysfnt{1}$
and $\sysfnt{2}$ contain the state to which $\mathsf{CS}_{\theta}$ is
to be applied.  The basic scheme is to start with two ancilla modes
$\sysfnt{3}$ and $\sysfnt{4}$ initialized with one photon each, apply
a linear optics transformation to modes
$\sysfnt{1},\sysfnt{2},\ldots,\sysfnt{k}$ with $k\geq 4$, measure all
but the first two of these modes and accept only a predetermined outcome,
say where one photon is detected in each of modes $\sysfnt{3}$ and
$\sysfnt{4}$ and none in the added modes.  Let $\hat u$ be the unitary
matrix associated with the linear optics transformation, with $u_{sr}$
the entries of $\hat u$.  The post-selected final state is determined
completely by the $4\times 4$ upper left submatrix $V$ of $\hat u$
with entries $V_{rs}=u_{sr}$ for $s,r\leq 4$.

It is necessary to consider the effects of the scheme on the initial
states $\kets{00}{12},\kets{01}{12},\kets{10}{12},\kets{11}{12}$.
Since photon number is conserved, we have, without renormalization:
\begin{eqnarray}
\kets{00}{12}&\rightarrow&\alpha_{0000}\kets{00}{12}
\label{eq:amps0}
\\
\kets{01}{12}&\rightarrow&\alpha_{0101}\kets{01}{12}+\alpha_{0110}\kets{10}{12}\\
\kets{10}{12}&\rightarrow&\alpha_{1010}\kets{10}{12}+\alpha_{1001}\kets{01}{12}\\
\kets{11}{12}&\rightarrow&\alpha_{1111}\kets{11}{12}+\alpha_{1120}\kets{20}{12}+\alpha_{1102}\kets{02}{12}
\label{eq:amps3}
\end{eqnarray}
To be successful, the amplitudes have to satisfy
\begin{eqnarray}
&\alpha_{0110}=\alpha_{1001}=\alpha_{1120}=\alpha_{1102}=0&
\label{eq:alpha0}\\
&\alpha_{1010}=\alpha_{0101}=\alpha_{0000}&
\label{eq:alpha1}\\
&\alpha_{1111}=e^{i\theta}\alpha_{0000}&
\label{eq:alpha2}
\end{eqnarray}
The amplitudes are polynomials of the coefficients of $V$.  For
example $\alpha_{0000}=v_{33}v_{44}+v_{34}v_{43}$.  More generally,
define $p_s = \sum_{r} v_{rs}\crtops{r}$.  If the initial state in
mode $\sysfnt{s}$ has $d_s$ photons, then the output state is given by
$P=\prod_s p_s^{d_s}$. Thus, $P$ is a polynomial of the
$\crtops{r}$. If $\beta$ is the coefficient of the monomial
$\prod_{t}\left(\crtops{t}\right)^{m_t}$ in $P$, then the output
amplitude for having $m_t$ photons in mode $\sysfnt{t}$ is given by
$\sqrt{\prod_t m_t!}\beta$. This shows that the amplitudes
$\alpha_{abcd}$ are polynomials of the $v_{rs}$.

The first step for constructing $\mathsf{CS}_{\theta}$ is to solve
Eqs.~\ref{eq:alpha0}--\ref{eq:alpha2}, which are polynomial identites
in the $v_{rs}$.  Before showing how to reduce the difficulty of doing
that, let us see how to proceed from there. Since there are $16$ free
complex variables, the solution will have a number of remaining free
variables that must be chosen to optimize the probability of success
($|\alpha_{0000}|^2)$ and to satisfy one more constraint: The solution
is an (explicit) matrix $V$ that needs to be extended to a unitary
matrix $\hat u$. This is possible if and only if the maximum singular
value (that is the square root of the maximum eigenvalue of $V^\dagger
V$) is at most one. The extension is not unique. One can set the first
four columns of $\hat u$ to the matrix with orthonormal columns
\begin{equation}
X =\left(\begin{array}{cc}
         V &                            \\
        (\idop-V^\dagger V)^{1/2}  &
         \end{array}\right),
\end{equation}
and then complete the last four columns with any orthonormal basis of
the orthogonal complement of the space spanned by the columns of $X$.
The maximum singular value constraint is needed to be able to compute
the square root in the expression for $X$. If some of the singular
values of $V$ are equal to one, then fewer than four additional
columns and rows can be used

The singular value constraint cannot be easily achieved using
algebraic methods. In principle, one can reparametrize the matrix $V$
to guarantee the constraint, for example by using the polar
decomposition and an Euler angle representation of unitary matrices.
In the case where $\mathsf{CS}_{\theta}$ is to be applied to the
``left'' modes of a pair of bosonic qubits, the singular value
constraint can be removed by exploiting a rescaling symmetry. Now
there are two additional modes to complete the bosonic qubits. The
total number of photons is always four. Let $V$ be a matrix whose
coefficients satisfy the identities for the $\alpha_{abcd}$. Let
$\lambda=\lambda(V)$ be the maximum singular value of $V$ and consider
the matrix
\begin{equation}
V_e = {1\over\lambda}\left(\begin{array}{cc}
              \idop&\mathbf{0}\\
              \mathbf{0}& V
            \end{array}\right),
\end{equation}
where the first two indeces are associated with the two other
(``right'') qubit modes. $V_e$ has maximum singular value $1$ and can
be extended to a unitary $\hat u_e$ as before.  The claim is that if
the resulting optics operation is applied to the pair of bosonic qubits with
the same post-selection procedure, it has the intended effect with
probability $1/\lambda^8$.  To see that this is true, first observe
that, $V'=\lambda V_e$ satisfies the polynomial equations obtained by
requiring that the operation works correctly for the pair of bosonic
qubits.  The amplitudes (as in Eqs.~\ref{eq:amps0}--\ref{eq:amps3})
that occur in these equations are polynomials which are either
homogenous linear in the coefficients of a given column of $V'$ or
independent of them.  This is because the input states have at most
one photon in each mode.  Because each input state under consideration
has exactly four photons, the amplitudes are all homogenous of degree
four in the coefficients of $V'$.  This implies that multiplying $V'$
by $\delta$ scales the amplitudes by $\delta^4$.  Since the equations
to be satisfied are homogenous linear in the amplitudes, every scalar
multiple of the matrix also satisfies the equations.

With the observation of the previous paragraph, instead of trying to
satisfy the singular value constraint, one can recalculate the
probability of success by dividing $V$'s probability of success by
$\lambda^8$ and optimize it.  Note that this works even if
$\lambda<1$.  In computer experiments using naive optimization methods
(see below), this usually led to solutions $V$ with $\lambda=1$ for
$\theta=180^\circ$ and $\theta=90^\circ$.

To simplify solving the equations for the $\alpha_{abcd}$, one can use
scaling symmetries to standardize $V$.  Since each of the
$\alpha_{abcd}$ is homogenous in the variables associated with any one
row or column of $V$, the equations of the form $\alpha_{abcd}=0$ are
satisfied for any rescaling of a row or column.  The non-zero
$\alpha_{abcd}$ are homogenous of degree one in the third and fourth
column (because of the presence of the helper photons at the input in
modes $\sysfnt{3}$ and $\sysfnt{4}$) and in the third and fourth row
(because of the post-conditioning on detecting exactly one photon in
each of modes $\sysfnt{3}$ and $\sysfnt{4}$). Because
Eqs.~\ref{eq:alpha1}--\ref{eq:alpha2} are homogenous linear in the
amplitudes, rescaling these rows or columns preserves the identities.
The non-zero $\alpha_{abcd}$ satisfy that they are of equal degree and
homogenous in the first column and (separately) in the first row. This
is due to the fact that when a photon is present in mode $\sysfnt{1}$
at the input, this is designed to be the case at the output too.  Thus
multiplying the first column by $\delta$ and the first row by
$1/\delta$ does not change the values. Similarly, this rescaling can
be used on the second column and row.

The scaling rules of the previous paragraph can be used to introduce
unconstrained scaling variables and standardize the entries of
$V$. For example, one can take $v_{13}=v_{24}=v_{33}=v_{44}=v_{43}=1$.
Note that this choice implies that solutions where any one of these
variables is $0$ are not easily found.  It may therefore be necessary
to try solving with some of the variables set to $0$.  For example,
the $\mathsf{CS}$ gate of~\cite{knill:qc2000e} after translation into
the form used here satisfies $v_{43}=0$.  I did not find any solutions
satisfying this constraint with better probability of success.

\begin{figure}
\[
\includegraphics[width=\textwidth]{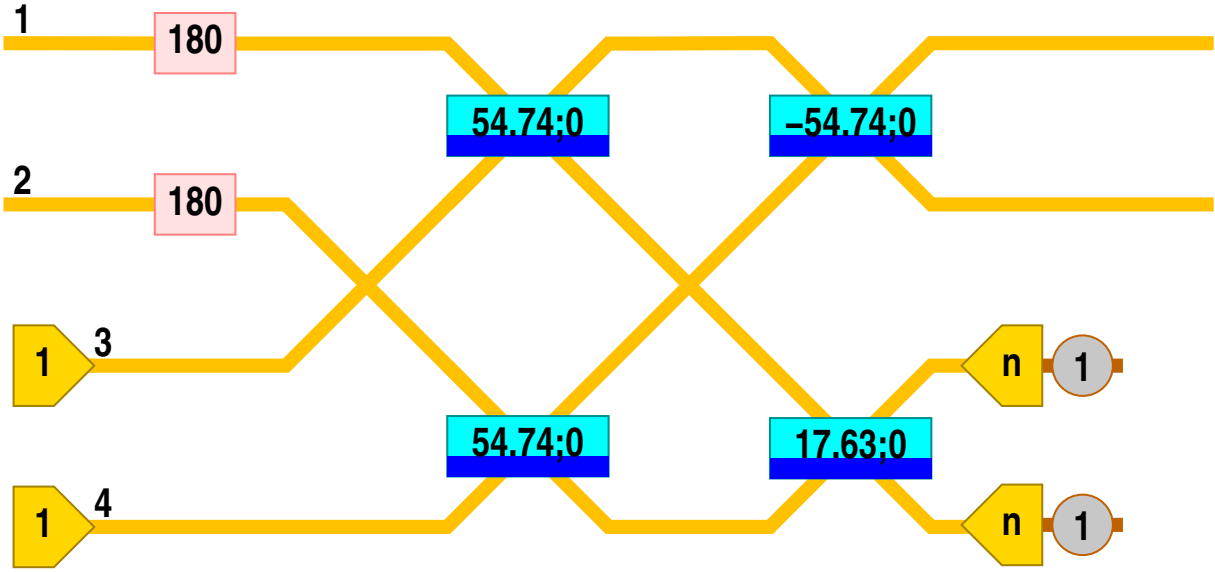}
\]
\caption{Optical network realizing $\mathsf{CS}_{180^\circ}$.
The notation is as explained in~\cite{knill:qc2000e}.}
\label{fig:cs180}
\end{figure}

Mathematica was used to solve the equations (any other computer
algebra system would do equally well). The strategy was to solve
linear equations first and then to simplify expressions.  Some
Mathematica notes are included verbatim in App.~\ref{app:math1} and
include formulas for the solution found. The solution could be
expressed in terms of the remaining variables of the last two columns
of $V$ and one additional variable.  After some experimentation, it
seemed that in all the best solutions, $v_{11}=v_{22}$. This was
exploited in the final version of the optimization procedure,
implemented in Matlab (the programs are available by request).
Briefly, the function to be optimized
takes as input the remaining free complex variables
($v_{14},v_{23},v_{34},l_1$), and a non-redundant subset of the
scaling variables. To avoid infinities, one can provide the logarithms of
the scaling variables as inputs. The scales can be
taken to be real since phases have no effect on the probability of
success. The function can then be optimized using random starting
points. With the optimization procedures provided by Matlab, it was found
useful to randomly perturb the point returned and repeat until the
solution no longer changes significantly. This procedure routinely
finds the same optimum. For $\theta=180^\circ$, it was possible
to guess the algebraic numbers that it converged to.
Here is a version of the matrix found, which turns out to be unitary:
\begin{equation}
  V_{180^\circ} = \left(\begin{array}{cccc}
     -1/3 & -\sqrt{2}\;/3 &  \sqrt{2}\;/3 &  2/3
       \\ 
     \sqrt{2}\;/3 &  -1/3 &  -2/3 &  \sqrt{2}\;/3
       \\ 
     -\sqrt{3 + \sqrt{6}}\;/3 & \sqrt{3 - \sqrt{6}}\;/3 & -\sqrt{(3 + \sqrt{6})/2}\;/3 &    \sqrt{1/6 - 1/(3\sqrt{6})}
       \\
     -\sqrt{3 - \sqrt{6}}\;/3 & -\sqrt{3 + \sqrt{6}}\;/3 & -\sqrt{1/6 - 1/(3\sqrt{6})} & -\sqrt{(3 + \sqrt{6})/2}\;/3
                  \end{array}\right)
\end{equation}
The probability of success is $2/27$.  The matrix can be
systematically decomposed into elementary beam splitter and phase
shift operators~\cite{reck:qc1994a}. An optical network realizing it
is shown in Fig.~\ref{fig:cs180}.  The implementation uses fewer
elements (4 beam splitters, 4 modes, 2 photon counters, probability of
success $2/27$) than the solution in~\cite{knill:qc2000e} (6 beam
splitters, 6 modes, 2 photon counters, 2 photo-detectors, probability
of success $1/16$). As before,
the counters must be able to distinguish between zero, one, or more
than one photons.

A matrix that can be extended to
obtain $\mathsf{CS}_{90^\circ}$ by post-selection was
also obtained:
\begin{equation}
V_{90^\circ}=\left(\begin{array}{@{}llll@{}}
  -0.3202 + 0.0418i & -0.2520 - 0.3226i &  0.2883  &          -0.1292 - 0.7221i\\
  -0.2520 - 0.3226i&  -0.3202 + 0.0418i&  -0.1292 - 0.7221i&   0.2883 \\
  -0.3216 + 0.7210i&  -0.1711 - 0.1725i&   0.2469          &   0.3322 + 0.3285i\\
  -0.1711 - 0.1725i&  -0.3216 + 0.7210i&   0.3322 + 0.3285i&   0.2469
  \end{array}\right)
\end{equation}
The probability of success for this solution is $1/19.37$.  A 
``nice'' beam splitter decomposition of this matrix was not found. Partly
this is due to the fact that because only two of the singular values
are (close to) one, at least two extra modes must be added for the
unitary completion.  The simplest method of decomposing a
$6\times 6$ unitary matrix normally requires $15$ beam splitters.

It is an open problem to determine whether the above solutions are
indeed optimal as is suggested by the results of the numerical
experiments.

\section{Bounds on conditional phase shifts?}

To obtain bounds on the probability of success of a phase-shift
gate implemented with helper photons, one can attempt to characterize
the states obtained in the output modes after tracing out the
helper modes. There is some choice of the initial state
of the modes that the gate is applied to. Assume
that this is a state obtained by applying linear optics
to prepared single photons. In this case,
the final state after a linear optics transformation
is given by 
\begin{equation}
\ket{\psi_f} = \prod_{k=1}^n (\alpha_{k1}\crtops{1}+\ldots+\alpha_{km}\crtops{m})\vacuum.
\end{equation}
The goal is to show that after tracing out modes $m'+1,\ldots,m$,
the state in the remaining modes is a mixture
of states of the form
\begin{equation}
\prod_{k=1}^n(\beta_{k0}+\beta_{k1}\crtops{1}+\ldots+\beta_{km'}\crtops{m'})\vacuum.
\end{equation}
In fact, this is the case if the final state before tracing out is also of
this form. To be explicit, add to the factors in the expression
for $\ket{\psi_f}$ any constant terms $\alpha_{k0}$ so that
\begin{equation}
\ket{\psi_f} = \prod_{k=1}^n (\alpha_{k0}+\alpha_{k1}\crtops{1}+\ldots+\alpha_{km}\crtops{m})\vacuum.
\end{equation}
First trace out mode $\sysfnt{m}$. Given an overcomplete
set of states $\kets{\gamma}{m}$, the state of
modes $\sysfnt{1}\ldots\sysfnt{m}$ is a mixture of the
(unnormalized) states $\bras{\gamma}{m}\ket{\psi_f}$.
Choosing as the set of states the coherent states and using
the fact that for these states
$\bras{\gamma}{m}\crtops{m}=\bras{\gamma}{m}\bar\gamma$,
the mixture consists of states of the form 
\begin{equation}
\prod_{k=1}^n(\alpha_{k0}+\bar\gamma\alpha_{km}+\alpha_{k1}\crtops{1}+\ldots+\alpha_{k(m-1)}\crtops{m-1})\vacuums{1\ldots(m-1)}.
\end{equation}
Iterating this procedure proves the desired result.

Consider the conditional sign-flip gate. With this gate and using a
few beam splitters, one can map the state $\ket{1100}$ to the state
${1\over \sqrt{2}}(\ket{1100}+\ket{0011}) =
{1\over\sqrt{2}}(\crtops{1}\crtops{2}+\crtops{3}\crtops{4})$, a
well-known entangled photon state. By the above, before post-selection
on a measurement of the other modes and with $n$ helper photons, the
state can be written as a mixture of products of linear expressions in
the creation operators.  To obtain a bound on the probability of
success, it suffices to obtain a bound for the overlap of (normalized)
such states with the Bell state.  Because the normalized overlap of
$(\crtops{1}+\crtops{3})(\crtops{2}+\crtops{4})$ with the Bell state
is $1/\sqrt{2}$, the bound on the probability of success thus obtained
can be no smaller than $1/2$.  It is clear that the probability of
success cannot be made equal to one: The polynomial $xy+uv$ associated
with the creation operators in the Bell state cannot be factored.

A problem suggested by the above is:

\noindent\textbf{Problem.} What is the maximum probability of success
for implementing $\mathsf{CS}_\theta$ using linear optics with at most
$k$ independently prepared helper photons and post-selection from
photon counters without feedback?

It was shown that for $\theta=180^\circ$, a probability of success of
one is not possible, but for $k\geq 2$, $1/13.5$ can be realized. A
variant of the problem asks the same question for conditional sign
shifts of two bosonic qubits (a four mode operation).  Other
directions for investigation are to determine what improvements are
possible if active linear optics operations can be used, or if initial
states such as prepared entangled photon pairs~\cite{pittman:qc2001a}
or photon number states like $\ket{2}$ are available.

\noindent{\bf Acknowledgements.} This work was partially supported by
the NSA and the DOE (contract W-7405-ENG-36).

\bibliographystyle{unsrt}
\bibliography{journalDefs,qc}

\appendix

\section{Mathematica Notes for Solving Eq.~\ref{eq:alpha0}--\ref{eq:alpha2}}
\label{app:math1}

The following gives a sequence of steps using Mathematica for solving
Eq.~\ref{eq:alpha0}--\ref{eq:alpha2}. 

\ignore{
perl -n -e 's/v([1-4])([1-4])/v$2$1/g; print;' ndgates.tex
}

\begin{verbatim}
(* The columns of V as polynomials of the creation operators
 * c1,c2,c3,c4  with the scaling simplifications. *)
p1 = (v11*c1+v21*c2+v31*c3+v41*c4); 
p2 = (v12*c1+v22*c2+v32*c3+v42*c4); 
p3 = (c1+v23*c2+c3+c4); 
p4 = (v14*c1+c2+v34*c3+c4); 

(* The amplitudes that occur in the equations to be solved are: *)

a0000 := Coefficient[p3*p4, c3*c4]; 
a1010 := Coefficient[p1*p3*p4, c3*c4*c1];
a1001 := Coefficient[p1*p3*p4, c3*c4*c2];
a0110 := Coefficient[p2*p3*p4, c3*c4*c1];
a0101 := Coefficient[p2*p3*p4, c3*c4*c2];
a1111 := Coefficient[p1*p2*p3*p4, c1*c2*c3*c4];
a1120 := Coefficient[p1*p2*p3*p4, c1^2*c3*c4];
a1102 := Coefficient[p1*p2*p3*p4, c2^2*c3*c4];

(* First solve a1001==0. 
 * a1001 is linear in the coefficients of p1. *)
p1x34 = Coefficient[a1001, v21];
p1x24 = Coefficient[a1001, v31];
p1x23 = Coefficient[a1001, v41];
(* Parametrize the linear solutions by hand, introducing l12,l13,l14: *)
rl1 = {v21->l12*p1x24 + l13*p1x23,
       v31->l12*(-p1x34) + l14*p1x23,
       v41->l13*(-p1x34) + l14*(-p1x24)}; 
p1 = p1/.rl1;

(* Similarly for a0110==0. *)
p1x34 = Coefficient[a0110, v12];
p1x24 = Coefficient[a0110, v32];
p1x23 = Coefficient[a0110, v42];
rl2 = {v12->l21*p1x24 + l23*p1x23,
       v32->l21*(-p1x34) + l24*p1x23,
       v42->l23*(-p1x34) + l24*(-p1x24)}; 
p2 = p2/.rl2;

(* Next solve a1010==a0101==a0000: 
 * Again this leads to linear equations in v11 and v22 respectively. *)
x1 = Coefficient[a1010,v11];
ap1010 = Simplify[a1010-x1*v11];
rl3 = {v11->(a0000-ap1010)/x1};
p1 = p1/.rl3;

x2 = Coefficient[a0101,v22];
ap0101 = Simplify[a0101-x2*v22];
rl4 = {v22->(a0000-ap0101)/x2};
p2 = p2/.rl4;

(* Now solve for a1120==0 and a1102==0.
 * This leads to quadratic equations in l12 and l21.
 * Explicitly: *)

ll11 = Coefficient[FullSimplify[a1120],l12*l21];
tl1120 = a1120-ll11*l12*l21;
ll01 = Coefficient[tl1120,l21];
ll10 = Coefficient[tl1120,l12];
ll00 = FullSimplify[tl1120-ll01*l21-ll10*l12];

lm11 = Coefficient[FullSimplify[a1102],l12*l21];
tl1102 = a1102-lm11*l12*l21;
lm01 = Coefficient[tl1102,l21];
lm10 = Coefficient[tl1102,l12];
lm00 = FullSimplify[tl1102-lm01*l21-lm10*l12];

xysols = Solve[x*y*ll11 + x*ll10 + y*ll01 + ll00 == 0 &&
      x*y*lm11 + x*lm10 + y*lm01 + lm00 == 0, {x,y}];
xy1 = FullSimplify[xysols[[1]]];

p1 = (p1/.{l12->x,l21->y})/.xy1;
p2 = (p2/.{l12->x,l21->y})/.xy1;

(*
 * One can now simplify the expressions by removing
 * redundant variables introduced earlier. p1's coefficients
 * are a function of Coefficient[p1,c4] and similarly for
 * p2. *)
lsimrule = {l13->(l1-l14*(1+v23))/(1+v34),
            l23->(l2-l24*(1+v14))/(1+v34)};
p1 = p1/.lsimrule;
p2 = p2/.lsimrule;
(* Check the identities by evaluating: 
 * Answers included after the expression:
FullSimplify[a0000]//InputForm
  1 + v34
FullSimplify[a0101]//InputForm
  1 + v34
FullSimplify[a1010]//InputForm
  1 + v34
FullSimplify[a0110]//InputForm
  0
FullSimplify[a1001]//InputForm
  0
FullSimplify[a1102]//InputForm
  0
FullSimplify[a1120]//InputForm
  0
FullSimplify[a1111]//InputForm
  (-1 - 4*l1*l2*v23^2*v14^2*(-1 + v34)*v34 + 
  v34*(-1 - 4*l1*l2*(-1 + v34) + v34 + v34^2 + 2*v23*(1 + v34)^2) + 
  v14*(-2*(1 + v34)^2 + v23*(-1 + v34)*(1 + v34*(2 + 8*l1*l2 + v34))))/
 ((1 + v34)*(-1 - (2 + v23)*v14 + v34 + v23*(2 + v14)*v34))
*)

(* Remaining identity: a1111 = ph*a0000, where ph is the desired phase. 
 * Note that a1111 is now a function of l1*l2, so solve for that. *)
l1rls = Solve[FullSimplify[a1111 == ph*a0000], {l1}];
l1rl = FullSimplify[l1rls[[1]]];
l1tl2 = l1*l2/.l1rl;

(*
 * Checking shows that p1-c1 and p2-c2 are multiples 
 * of l1 and l2 respectively. Exploit that to express
 * the coefficients of p1 and p2: *)
dp1c1 = FullSimplify[(Coefficient[p1,c1]-1)/l1];
dp1c2 = FullSimplify[Coefficient[p1,c2]/l1];
dp1c3 = FullSimplify[Coefficient[p1,c3]/l1];
dp1c4 = FullSimplify[Coefficient[p1,c4]/l1];

dp2c1 = FullSimplify[Coefficient[p2,c1]/l2];
dp2c2 = FullSimplify[(Coefficient[p2,c2]-1)/l2];
dp2c3 = FullSimplify[Coefficient[p2,c3]/l2];
dp2c4 = FullSimplify[Coefficient[p2,c4]/l2];

(* The transpose of V is now given by:
*)
vmat = {{dp1c1,dp1c2,dp1c3,dp1c4}*l1+{1,0,0,0},
        {dp2c1,dp2c2,dp2c3,dp2c4}*l2+{0,1,0,0},
        {1,v23, 1, 1},
        {v14,1,v34,1}};
(* Formulas:
dp1c1//InputForm
(1 + v14 + v23*v14 - 2*v14^2 - v23*v14^2 - 2*v34 + 2*v23*v34 - 4*v14*v34 + 
  4*v23*v14*v34 - 2*v14^2*v34 + 2*v23*v14^2*v34 + v34^2 + 2*v23*v34^2 - 
  v14*v34^2 - v23*v14*v34^2 - v23*v14^2*v34^2 + 
  (1 + v14)*Sqrt[v14^2*(v23*(-1 + v34)^2 + 2*(1 + v34))^2 + 
     2*v14*(2*(-1 + v34)^2*(1 + v34) + 2*v23^2*(-1 + v34)^2*v34*(1 + v34) + 
       v23*(1 - 18*v34^2 + v34^4)) + (1 + v34*(-2 + v34 + 2*v23*(1 + v34)))^
      2])/(2*(1 + v34)*(-1 - (2 + v23)*v14 + v34 + v23*(2 + v14)*v34))
dp1c2//InputForm
(-1 + v23 - 2*v14 + v23*v14 + v23^2*v14 + 2*v34 + 4*v23*v34 + 2*v23^2*v34 - 
  2*v14*v34 - 4*v23*v14*v34 - 2*v23^2*v14*v34 - v34^2 - v23*v34^2 + 
  2*v23^2*v34^2 - v23*v14*v34^2 + v23^2*v14*v34^2 + 
  (1 + v23)*Sqrt[v14^2*(v23*(-1 + v34)^2 + 2*(1 + v34))^2 + 
     2*v14*(2*(-1 + v34)^2*(1 + v34) + 2*v23^2*(-1 + v34)^2*v34*(1 + v34) + 
       v23*(1 - 18*v34^2 + v34^4)) + (1 + v34*(-2 + v34 + 2*v23*(1 + v34)))^
      2])/(2*(1 + v34)*(-1 - (2 + v23)*v14 + v34 + v23*(2 + v14)*v34))
dp1c3//InputForm
((-1 + v34)*(1 + (2 + v23)*v14 + v34 + v23*(2 + v14)*v34) - 
  Sqrt[v14^2*(v23*(-1 + v34)^2 + 2*(1 + v34))^2 + 
    2*v14*(2*(-1 + v34)^2*(1 + v34) + 2*v23^2*(-1 + v34)^2*v34*(1 + v34) + 
      v23*(1 - 18*v34^2 + v34^4)) + (1 + v34*(-2 + v34 + 2*v23*(1 + v34)))^
     2])/(2*(-1 - (2 + v23)*v14 + v34 + v23*(2 + v14)*v34))
dp1c4//InputForm
 -1
dp2c1//InputForm
(1 + v14 + v23*v14 - 2*v14^2 - v23*v14^2 - 2*v34 + 2*v23*v34 - 4*v14*v34 + 
  4*v23*v14*v34 - 2*v14^2*v34 + 2*v23*v14^2*v34 + v34^2 + 2*v23*v34^2 - 
  v14*v34^2 - v23*v14*v34^2 - v23*v14^2*v34^2 - 
  (1 + v14)*Sqrt[v14^2*(v23*(-1 + v34)^2 + 2*(1 + v34))^2 + 
     2*v14*(2*(-1 + v34)^2*(1 + v34) + 2*v23^2*(-1 + v34)^2*v34*(1 + v34) + 
       v23*(1 - 18*v34^2 + v34^4)) + (1 + v34*(-2 + v34 + 2*v23*(1 + v34)))^
      2])/(2*(1 + v34)*(-1 - (2 + v23)*v14 + v34 + v23*(2 + v14)*v34))
dp2c2//InputForm
(-1 + v23 - 2*v14 + v23*v14 + v23^2*v14 + 2*v34 + 4*v23*v34 + 2*v23^2*v34 - 
  2*v14*v34 - 4*v23*v14*v34 - 2*v23^2*v14*v34 - v34^2 - v23*v34^2 + 
  2*v23^2*v34^2 - v23*v14*v34^2 + v23^2*v14*v34^2 - 
  (1 + v23)*Sqrt[v14^2*(v23*(-1 + v34)^2 + 2*(1 + v34))^2 + 
     2*v14*(2*(-1 + v34)^2*(1 + v34) + 2*v23^2*(-1 + v34)^2*v34*(1 + v34) + 
       v23*(1 - 18*v34^2 + v34^4)) + (1 + v34*(-2 + v34 + 2*v23*(1 + v34)))^
      2])/(2*(1 + v34)*(-1 - (2 + v23)*v14 + v34 + v23*(2 + v14)*v34))
dp2c3//InputForm
((-1 + v34)*(1 + (2 + v23)*v14 + v34 + v23*(2 + v14)*v34) + 
  Sqrt[v14^2*(v23*(-1 + v34)^2 + 2*(1 + v34))^2 + 
    2*v14*(2*(-1 + v34)^2*(1 + v34) + 2*v23^2*(-1 + v34)^2*v34*(1 + v34) + 
      v23*(1 - 18*v34^2 + v34^4)) + (1 + v34*(-2 + v34 + 2*v23*(1 + v34)))^
     2])/(2*(-1 - (2 + v23)*v14 + v34 + v23*(2 + v14)*v34))
dp2c4//InputForm
-1
* And l1*l2 == 
l1tl2//InputForm
  -((-1 + ph)*(1 + v34)^2*(-1 - (2 + v23)*v14 + v34 + v23*(2 + v14)*v34))/
   (4*(-1 + v23*v14)^2*(-1 + v34)*v34)
*)



\end{verbatim}

\end{document}